\documentclass[%
reprint,
superscriptaddress,
showpacs,preprintnumbers,showkeys,
amsmath,amssymb,
aps,
]{revtex4-1}

\usepackage{graphicx}
\usepackage{dcolumn}
\usepackage{bm}

\begin{document}


\title{Effect of Cr spacer on structural and magnetic properties of Fe/Gd multilayers}

\author{A.~B.~Drovosekov}
\email{drovosekov@kapitza.ras.ru}
\affiliation {P.L.Kapitza Institute for Physical Problems, RAS, Moscow 119334, Russia}

\author{M.~V.~Ryabukhina}
\email{ryabukhina@imp.uran.ru}
\affiliation{M.N.Mikheev Institute of Metal Physics, UB RAS, Ekaterinburg 620990, Russia}

\author{D.~I.~Kholin}
\author{N.~M.~Kreines}
\affiliation {P.L.Kapitza Institute for Physical Problems, RAS, Moscow 119334, Russia}

\author{E.~A.~Manuilovich}
\affiliation {P.L.Kapitza Institute for Physical Problems, RAS, Moscow 119334, Russia}
\affiliation {Moscow Institute of Physics and Technology, Dolgoprudny 141701, Russia}

\author{A.~O.~Savitsky}
\affiliation {P.L.Kapitza Institute for Physical Problems, RAS, Moscow 119334, Russia}
\affiliation {Institute of Solid State Physics, RAS, Chernogolovka 142432, Russia}

\author{E.~A.~Kravtsov}
\affiliation{M.N.Mikheev Institute of Metal Physics, UB RAS, Ekaterinburg 620990, Russia}
\affiliation{Ural Federal University, Ekaterinburg 620002, Russia}

\author{V.~V.~Proglyado}
\affiliation{M.N.Mikheev Institute of Metal Physics, UB RAS, Ekaterinburg 620990, Russia}

\author{V.~V.~Ustinov}
\affiliation{M.N.Mikheev Institute of Metal Physics, UB RAS, Ekaterinburg 620990, Russia}
\affiliation{Ural Federal University, Ekaterinburg 620002, Russia}

\author{T.~Keller}
\author{Yu.~N.~Khaydukov}
\affiliation{Max-Planck-Institut f\"ur Festk\"orperforschung, Heisenbergstra{\ss}e 1, D-70569 Stuttgart, Germany}
\affiliation{Max Planck Society Outstation at the Heinz Maier-Leibnitz Zentrum (MLZ), D-85748 Garching, Germany}

\author{Y.~Choi}
\author{D.~Haskel}
\affiliation{Advanced Photon Source, Argonne National Laboratory, Argonne, Illinois 60439, USA}

\begin{abstract}
In this work we analyse the role of a thin Cr spacer between Fe and Gd layers on structure and magnetic properties of a [Fe(35\,\AA)/Cr($t_\mathrm{Cr}$)/Gd(50\,\AA)/Cr($t_\mathrm{Cr}$)]$_{12}$ superlattice. Samples without the Cr spacer ($t_\mathrm{Cr}=0$) and with a thin $t_\mathrm{Cr}=4$~\AA{} are investigated using X-ray diffraction, polarized neutron and resonance X-ray magnetic reflectometry, static magnetometery, magneto-optical Kerr effect and ferromagnetic resonance techniques. Magnetic properties are studied experimentally in a wide temperature range $4-300$~K and analysed theoretically using numerical simulation on the basis of the mean-field model. We show that a reasonable agreement with the experimental data can be obtained considering temperature dependence of the effective field parameter in gadolinium layers. The analysis of the experimental data shows that besides a strong reduction of the antiferromagnetic coupling between Fe and Gd, the introduction of Cr spacers into Fe/Gd superlattice leads to modification of both structural and magnetic characteristics of the ferromagnetic layers.
\end{abstract}

\pacs{68.65.Ac, 75.70.Cn, 61.05.-a}
\keywords{Fe/Gd multilayers, magnetic properties, interlayer coupling, mean-field model}

\maketitle


\section{Introduction}

In the last decades investigations of magnetic multilayers have attracted attention due to a variety of unique magnetic properties and unusual magnetic phenomena. Layered structures based on transition (3d) and rare-earth (4f) ferromagnetic (FM) metals, like Fe/Gd, are model ferrimagnet systems demonstrating a rich magnetic phase diagram with complex types of magnetic ordering \cite{Cam2015,Cam1993,Cam1988,Cam1989,LePage1990}. The magnetic state in the Fe/Gd multilayer is governed by several competing factors: a strong antiferromagnetic (AFM) coupling at Fe-Gd interfaces, enhancement of Gd magnetic moment in the interfacial region near Fe, Zeeman interaction with the external field \cite{Cam2006}. Camley \textsl{et~al.} calculated $H-T$ phase diagrams for Fe/Gd systems, showing the possibility of so-called Fe-aligned, Gd-aligned, and twisted magnetic phases \cite{Cam1988,LePage1990,Cam1987}. Experimental realization of such phases was clearly demonstrated by the resonant X-ray magnetic reflectometry technique in a number of works \cite{Ishi1999,Hos2002,Hask2001,Choi2004,Kra2009}.

Recently a new rise of interest to 3d/4f multilayers is caused by observations of magnetic skyrmion states in Fe/Gd system \cite{Mon2017,Mont2017}. Studies of magnetization dynamics in 3d/4f systems attract attention due to a recent idea to use such materials for realization of ultrafast magnetic switching, promising for potential applications in magnetic storage devices \cite{Man2014}. In particular, Fe/Gd multilayers and amorpous alloys are the systems of this sort \cite{Chim2015,Xu2016}.

Combined 3d-4f layered structures with mediating non-magnetic spacers have been recently considered as systems for realization of a high magnetic moment at room temperature \cite{Scheun2016}. AFM chromium was proposed as a spacer that could potentially initiate a strong FM coupling between rare-earth and transition metal layers, leading to an enhancement of the magnetic moment and high Curie temperature in such combined systems \cite{Sany2010}. However, experimental results performed on Fe/Cr/Gd \cite{Sany2010,Strom2011,Drov2015,Sun2018} and FeCo/Cr/Gd \cite{Ward2013} systems have not shown the desired moment improvement. Moreover, the FM layers demonstrate reduced values of the saturation magnetization which can be caused by imperfections of interfaces and crystal structure in the superlattice \cite{Ward2013,Scheun2012}.

In previous work \cite{Kra2017}, we demonstrated that introduction of the Cr spacer between FM layers in the Fe/Gd superlattice initiates a structural modification of Gd layers. Formation of the fcc crystallographic phase within the Gd layers in addition to the hcp phase seems to be one of the factors leading to reduced Gd magnetization in the Fe/Cr/Gd structure.

In this work, we focus on the effect of the Cr spacer on magnetic characteristics of the system. To obtain detailed information about modification of the magnetic parameters, we perform complex investigations of the static magnetization, magnetic resonance, and magneto-optical properties of the Fe/Gd and Fe/Cr/Gd superlattices. To probe magnetization depth profiles in the samples, we perform complementary measurements of polarized neutron reflectometry and resonant x-ray magnetic reflectometry which are known as the most powerful techniques to precisely resolve (at subnanometer scale) inhomogeneous magnetization density within magnetic heterostructures \cite{Zhu2015,Duf1993,Hahn1995,McGrath1996,Roy2005}.

To obtain magnetic parameters of the system, the experimental data are compared with numerical simulations on the basis of the mean-field approach. The mean-field model is a method which is commonly used to analyse the complex magnetic states in Fe/Gd systems \cite{Cam1987,Drov2017}. Recently the similar approach was used to simulate magnetization reversal in Py/Gd \cite{Lapa2017} and Ni/Gd \cite{Higgs2016} heterostructures. In spite of its simplicity, the mean-field model predicts all the main features of the considered systems. However, quantitative agreement with experiment is under question. Detailed magnetization data obtained in a wide range of temperatures and magnetic fields are described only qualitatively in the frame of the effective field model \cite{Tak1992}. The temperature dependence of magnetization in Gd layers was reported to be close to linear \cite{Hos2002jpcm} which contradicts the standard mean-field theory.

In previous work \cite{Drov2017}, we analysed magnetic properties of a Fe/Gd superlattice in the frame of modified mean-field model with temperature dependent effective field constant. The proposed approach was proved to provide good description of both static magnetization and ferromagnetic resonance data obtained experimentally in a wide $4-300$~K temperature range. In view of this, it would be interesting to perform further investigations of the applicability of the proposed approach to analysis of layered systems of this sort, such as Fe/Cr/Gd.

Indeed, here we show that for both Fe/Gd and Fe/Cr/Gd structures a reasonable agreement with the experimental data can be obtained considering temperature dependence of the effective field parameter in gadolinium layers \cite{Drov2017}. The analysis of the experimental data shows that the introduction of Cr spacers into Fe/Gd superlattice leads to a strong reduction of the AFM coupling between Fe and Gd layers and to modification of both structural and magnetic characteristics of the FM layers.

\section{Samples and experimental techniques}

The multilayer structures, [Fe($t_\mathrm{Fe}$)/Gd($t_\mathrm{Gd}$)]$_{12}$ and [Fe($t_\mathrm{Fe}$)/Cr($t_\mathrm{Cr}$)/Gd($t_\mathrm{Gd}$)/Cr($t_\mathrm{Cr}$)]$_{12}$, with nominal layer thicknesses $t_\mathrm{Fe}\approx35$~\AA{}, $t_\mathrm{Gd}\approx50$~\AA{} and $t_\mathrm{Cr}\approx4$~\AA{} were prepared using high vacuum magnetron sputtering technique. The superlattices were deposited on glass and Si(100) substrates with 50~\AA{} thick chromium buffer layer. To prevent oxidation, a 30~\AA{} chromium cap layer was deposited on the top of the structure. For convenience, in this work we will refer to the superlattices with and without Cr spacers as ``Fe/Cr/Gd'' and ``Fe/Gd'' respectively. Samples prepared on different substrates proved to demonstrate identical structural and magnetic characteristics.

The structural characterization was performed by conventional X-ray diffraction (XRD), grazing incidence X-ray diffraction (GIXRD) and X-ray reflectometry (XRR). The measurements were carried out on a laboratory Empyrean PANalytical diffractometer using either Cu$K_\alpha$ or Co$K_\alpha$ radiation.

Static magnetization was investigated in $4-300$~K temperature range in magnetic fields up to 50~kOe, using a conventional SQUID magnetometer Quantum Design MPMS. Magnetic properties of the substrate were measured separately and its contribution was subtracted from the total magnetic moment of the samples.

Ferromagnetic resonance (FMR) was studied using a laboratory developed transmission type spectrometer in the range of frequencies $7-37$~GHz at temperatures $4-300$~K in magnetic fields up to 10~kOe.

Longitudinal magneto-optical Kerr effect (MOKE) studies of the surface magnetization were performed in $4-300$~K temperature range in magnetic fields up to 10~kOe using a 635~nm semiconductor laser.

The magnetization distribution in the superlattices was determined using the resonance X-ray magnetic reflectivity (RXMR) and polarized neutron reflectometry (PNR) experiments at $T=15$~K in magnetic field $H=500$~Oe.

RXMR measurements were performed at undulator beamline 4ID-D of the Advanced Photon Source at Argonne National Laboratory \cite{Lang1995}. Magnetic reflectivity scans were done at the Gd $L_2$ resonance $2p_{1/2}\rightarrow5d$ excitation with photon energy $E=7929$~eV. The magnetic reflectivity was measured as the difference between reflected intensities of the circularly polarized light for  two opposite helicities ($R^+-R^-$).

PNR experiment was conducted on the angle-dispersive reflectometer NREX at the research reactor FRM~II the Heinz Maier-Leibnitz Zentrum in Garching, Germany. The NREX measurements were done in standard $\theta-2\theta$ geometry with constant neutron wavelength of $4.26\pm0.06$~\AA{} and polarization 99.99\%. The polarization of the reflected beam was analyzed by a polarization analyzer with efficiency 98\%.

In all the experiments, the external magnetic field was applied in the film plane.

\section{Mean-field model}

To define magnetic parameters of the samples, the experimental data were compared with calculations based on the mean-field approach. The general idea of calculation procedure is similar to that described in \cite{Cam1993} and more details can be find in our work \cite{Drov2017}. Due to a high $T_\mathrm{C}$ and a large exchange stiffness of Fe layers, they are considered as homogeneously magnetized up to saturation value $M_\mathrm{Fe}$ at temperatures under study. To model the magnetization distribution in Gd layers, they are divided into 16 sublayers with thickness $a\approx3$~\AA{} (formally corresponding to the distance between hexagonal atomic planes in hcp Gd). Thus the total superlattice is divided into $12\times17=204$ elementary sublayers and we come to the problem to find the equilibrium magnetization in each of them. This problem can be solved using an iteration method. Starting from some initial distribution of magnetization $\mathbf{M}_i$, where $i$ is the index of sublayer, we may find the effective field $\mathbf{H}_i$ which acts on the spins in each sublayer. This effective field is the sum of the exchange field and the external field $\mathbf{H}$. To calculate the total exchange field acting on the spin in layer $i$, we must consider separately the contributions from the spins in the same layer $i$ and from the spins in neighbouring layers $i\pm1$. Thus, for the spins inside Gd layers we can write
\begin{equation}
\mathbf{H}_i = \mathbf{H} + \lambda [ \zeta \mathbf{M}_{i+1} + \zeta \mathbf{M}_{i-1} + (1-2\zeta) \mathbf{M}_i ],
\end{equation}
where $\lambda$ is the mean-field parameter of Gd and $\zeta$ characterizes the relative contribution of the neighbouring Gd sublayers in the total exchange field. In case of ideal crystal structure the parameter $\zeta$ can be treated as the fraction of nearest neighbour atoms in $i\pm1$ atomic layers, $z_{i\pm1}$, in the total number of nearest neighbours $z$, i.e. $\zeta=z_{i\pm1}/z$. On the other hand, the parameter $\zeta$ is directly connected with exchange stiffness $A$ of the Gd layer by relation
\begin{equation}
\label{ExchStiff}
A = \frac{1}{2} \zeta \lambda M^2 a^2.
\end{equation}
To find the exchange fields at Fe-Gd interfaces, we consider the Fe-Gd interlayer coupling energy per unit area in the form
\begin{equation}
E = -J\frac{(\mathbf{M}_i \mathbf{M}_{i+1})}{M_\mathrm{Fe} M_\mathrm{Gd}},
\end{equation}
where $M_\mathrm{Fe}$ and $M_\mathrm{Gd}$ are saturation magnetization for Fe and Gd respectively and $J$ is coupling constant. Here the indexes $i$ and $i+1$ are related to interfacial layers Fe and Gd. The corresponding exchange fields at the Fe-Gd interface are defined by
\begin{equation}
\mathbf{H}_i^\textrm{Fe-Gd} = - \frac{1}{t_i} \frac{\partial E}{\partial \mathbf{M}_i},
\end{equation}
where $t_i$ is thickness of layer $i$.

The first step of the considered iterative procedure is to find the equilibrium directions of vectors $\mathbf{M}_i$ which are defined by condition $\mathbf{M}_i \parallel \mathbf{H}_i$. As a second step, we must find the absolute values of $\mathbf{M}_i$. At this step we need to calculate only the magnetization for Gd sublayers because we neglect the temperature changes of Fe magnetization. We perform this calculation using the mean-field approach:
\begin{equation}
M_i = M_\mathrm{Gd} B_{S} \left( \frac{\mu H_i}{k_\mathrm{B} T} \right),
\end{equation}
where $B_{S}$ is the Brillouin function for Gd spin $S=7/2$, $\mu=7.5\mu_\mathrm{B}$ is the magnetic moment of Gd ion, $\mu_\mathrm{B}$ is Bohr magneton and $k_\mathrm{B}$ is Boltzman constant.

When the new $\mathbf{M}_i$ values are found, we return back to the first step and the procedure is repeated until the stationary self-consisted solution is found. The resulting total magnetic moment per unit area of the superlattice is defined by the expression:
\begin{equation}
\label{m}
m = \sum_{i} t_i M_i^\parallel,
\end{equation}
where $M_i^\parallel$ is magnetization component in the field direction.

After the calculation of the static magnetization distribution, we can analyse magnetic resonance properties of the system. Magnetization dynamics is described by Landau-Lifshitz equations (LLE) with relaxation terms ($\mathbf{R}_i$):
\begin{equation}
\label{L-L}
\frac{\partial \mathbf{M}_i}{\partial t}  = -\gamma_i [\mathbf{M}_i \times (\mathbf{H}_i - 4\pi \mathbf{M}_i^z)] + \mathbf{R}_i,
\end{equation}
where $\gamma$ is gyromagnetic ratio. Here, besides the external and exchange effective fields, we must take into account an additional demagnetization field $4\pi \mathbf{M}_i^z$ due to the presence of dynamical magnetization component $\mathbf{M}_i^z$ perpendicular to the film plane.

The FMR frequencies are defined as eigenfrequencies of linearised system \eqref{L-L}. Following our previous work \cite{Drov2017}, we restricted ourselves by considering only one period of the superlattice. Such approach proved to give sufficiently good approximation of the experimental spectra. At the same time, to achieve better agreement with the experiment, we considered the non-local dissipative term in equations (5) written in continual form as
\begin{equation}
\label{Damp}
\mathbf{R} = -A^* M_\mathrm{Gd} [\mathbf{m} \times \nabla^2 \frac{\partial \mathbf{m}}{\partial t}],
\end{equation}
where $\mathbf{m}$ is a unit vector in the direction of Gd magnetization, $A^*$ is a constant. This term provides extra suppression of the high-order spin-wave modes in Gd layer \cite{Drov2017}.

\begin{figure}[t]
\includegraphics[width=.85\columnwidth]{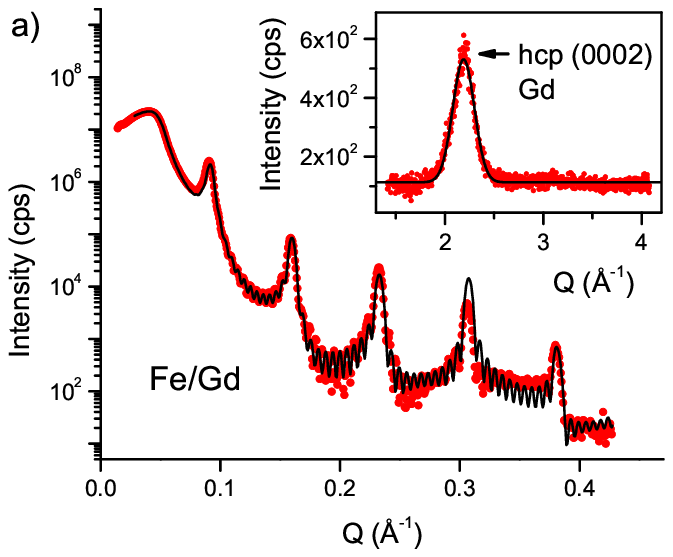}
\vskip 1mm
\includegraphics[width=.85\columnwidth]{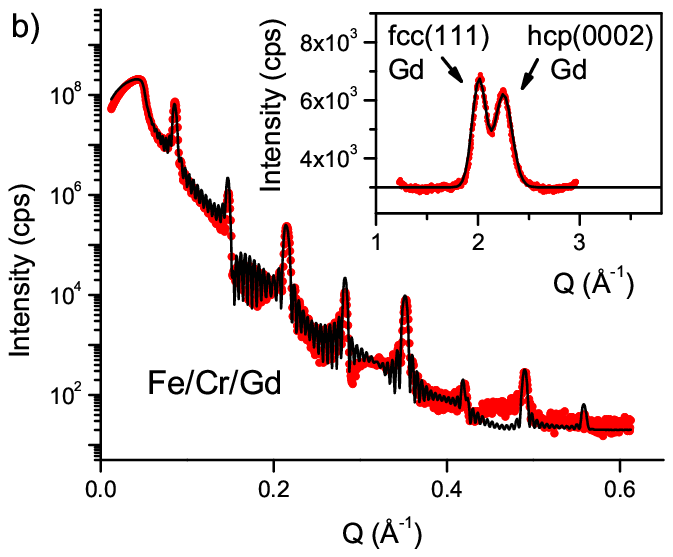}
\caption{X-ray reflectivity of the studied Fe/Gd (a) and Fe/Cr/Gd (b) multilayers. Points are the experimental data, lines demonstrate their approximation. The insets in (a) and (b) show the corresponding GIXRD patterns.}
\label{Xray}
\end{figure}

\section{Results}

\subsection{X-ray data}

Experimental XRR spectra for the studied samples are presented in Fig.~\ref{Xray}. The structural refinement results show that the samples have well-defined layered structure with interfacial root mean square roughness of about 1--2 atomic monolayers. The thicknesses of different layers in the superlattices extracted from XRR data are close to their nominal values ($t_\mathrm{Fe}=33\pm1$~\AA{}, $t_\mathrm{Gd}=48\pm2$~\AA{} and $t_\mathrm{Cr}=5\pm1$~\AA{}).

The crystal structure of the superlattices was investigated using XRD and GIXRD at fixed incident angle $\omega=3^\circ$. The average size of the hcp Gd crystallites in the studied Fe/Cr/Gd superlattices, which was estimated using the halfwidth of (0002) hcp Gd Bragg reflection and Debye-Scherrer equation, is about 20~\AA. The insets in Fig.~\ref{Xray} show the experimental GIXRD patterns. For the sample Fe/Gd the spectrum demonstrates only one very broad peak corresponding to (0002) hcp Gd reflection. We detected no signal from Fe which means that Fe layers are likely to be in amorphous state. For the Fe/Cr/Gd sample (Fig.~\ref{Xray}b), the GIXRD spectrum indicates the presence of different types of crystallites in Gd layers. Besides the (0002) hcp Gd reflection, the spectrum demonstrates additional peak corresponding to (111) fcc Gd reflection. This result is in accordance with previous work \cite{Kra2017} where the same effect of Cr spacer was observed for Fe/Cr/Gd superlattices prepared on Si substrates. Thus, thin Cr spacer between Fe and Gd layers significantly modifies the structural properties of Gd.

\subsection{Static magnetization}

\begin{figure}[t]
\includegraphics[width=.85\columnwidth]{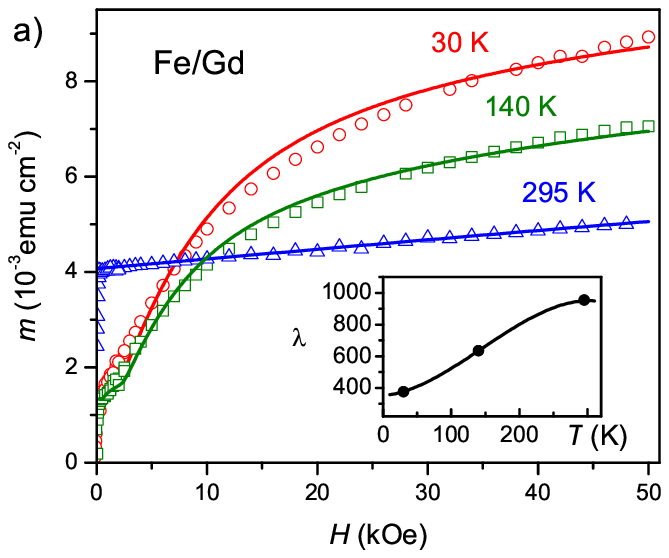}
\vskip 1mm
\includegraphics[width=.85\columnwidth]{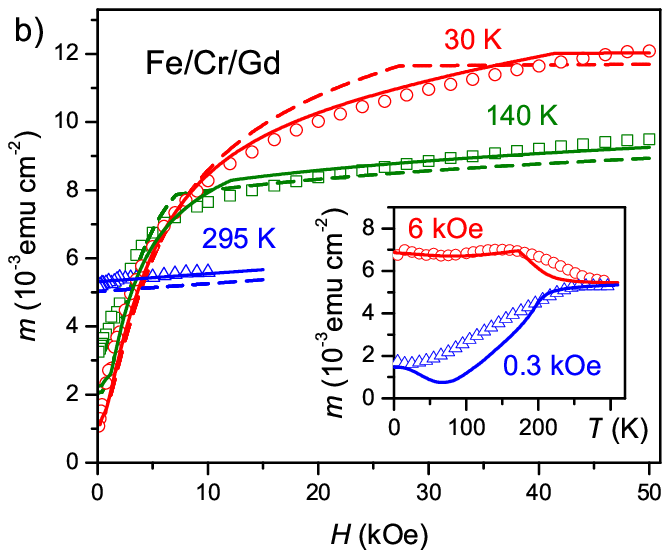}
\caption{(a) Experimental magnetization curves at $T=30$, 140 and 295~K (points) and their best fit within the mean-field model (lines) for the Fe/Gd sample. The inset shows temperature dependence of the mean-field parameter $\lambda(T)$ obtained in \cite{Drov2017}. (b) Magnetization curves at $T=30$, 140 and 295~K for the sample Fe/Cr/Gd. Points are the experimental data, dashed and solid lines are their mean-field approximation with different set of parameters (fit 1 and fit 2 respectively, see Table~\ref{Table}). The inset demonstrates magnetic moment per unit area as a function of temperature at $H=0.3$ and 6~kOe.}
\label{MH}
\end{figure}

Fig.~\ref{MH} shows experimental magnetization curves $m(T)$ at different temperatures and the result of their approximation within the mean field model with different sets of parameters shown in Tab.~\ref{Table}. The magnetization curves below $\sim200$~K have essentially non-linear form with smooth approach to saturation, indicating the twisted state in Gd layers.

\begin{table}[b]
\caption{\label{Table}
Mean-field model parameters for samples Fe/Gd and Fe/Cr/Gd.
}
\begin{ruledtabular}
\begin{tabular}{cccc}
& Fe/Gd&\multicolumn{2}{c}{Fe/Cr/Gd}\\
& (see \cite{Drov2017})&fit 1&fit 2\\
\colrule
$M_\mathrm{Fe}$ (emu\,$\cdot$\,cm$^{-3}$) & 1270 & 1270 & 1350\\
$M_\mathrm{Gd}$ (emu\,$\cdot$\,cm$^{-3}$) & 1150 & 1150 & 1150\\
$J$ ~ (erg\,$\cdot$\,cm$^{-2}$) & --39 & --2.0 & --2.5\\
$\zeta$ & 0.33 & 0.33 & 0.25\
\end{tabular}
\end{ruledtabular}
\end{table}

The fitting parameters for the sample Fe/Gd were obtained in \cite{Drov2017}. It was shown that much better fit of $m(H)$ curves can be obtained taking into account temperature dependence of the effective field parameter $\lambda$ in Gd layers (see the inset in Fig.~\ref{MH}a). To achieve the best approximation of the experimental $m(T)$ curves, in the work \cite{Drov2017} we considered polynomials of different order for the $\lambda(T)$ dependence. As a result we obtained reasonably good agreement with experiment using a third order polynomial:
\begin{equation}
\label{lambda}
\lambda(T) \approx 800 + 505 \tau - 255 \tau^2 - 310 \tau^3,
\end{equation}
where $\tau=(T-T_\mathrm{C})/T_\mathrm{C}$ with gadolinium Curie temperature $T_\mathrm{C}\approx200$~K.

In this work we used the obtained dependence $\lambda(T)$, Eq.~\eqref{lambda}, to analyse the magnetic properties of the Fe/Cr/Gd sample. Comparing the magnetization curves for samples Fe/Gd and Fe/Cr/Gd (Fig.~\ref{MH}), it is obvious that the insertion of the Cr spacer between Fe and Gd layers leads to significant increase of magnetic susceptibility of the system. This effect is clearly due to a strong reduction of the AFM interlayer coupling at Fe-Gd interface. Taking into account this argumentation, we tried to fit the experimental $m(H)$ curves for the Fe/Cr/Gd sample varying only the interlayer coupling parameter $J$, while other parameters of the system were equal to those for Fe/Gd sample (fit~1, Tab.~\ref{Table}). The result of such procedure is shown in Fig.~\ref{MH}b by dashed lines. Surprisingly, this simple approach allows to achieve reasonable qualitative agreement with the experimental data. On the other hand, there is a certain quantitative discrepancy between experimental and calculated curves $m(H)$. First, we notice that the experimental dependencies demonstrate larger saturation magnetization at all temperatures which can be due to increased magnetization of the Fe layers in the Fe/Cr/Gd sample. Second, the experimental curves $m(H)$ at low temperatures show smoother approach to saturation. This effect can be ascribed to smaller exchange stiffness of Gd layers in the Fe/Cr/Gd sample.


\begin{figure}[t]
\includegraphics[width=.85\columnwidth]{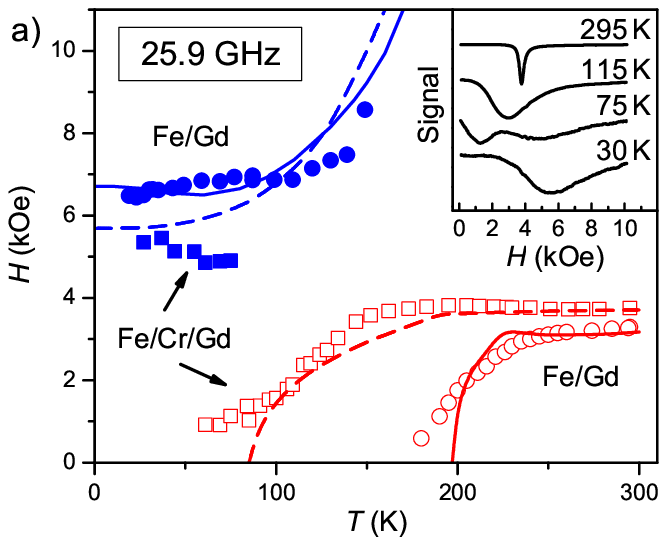}
\vskip 1mm
\includegraphics[width=.85\columnwidth]{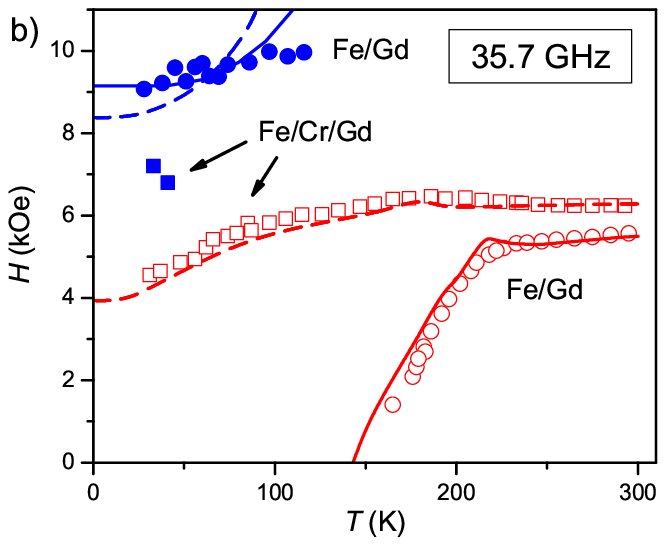}
\caption{Resonance field as a function of temperature for two samples at $f=25.9$~GHz (a) and $f=35.7$~GHz (b). Points are the experimental data, lines are the result of modelling. Inset in the graph (a) shows examples of resonance signal for sample Fe/Cr/Gd at different temperatures.}
\label{Hres}
\end{figure}

\begin{figure}[t]
\centering
\includegraphics[width=.85\columnwidth]{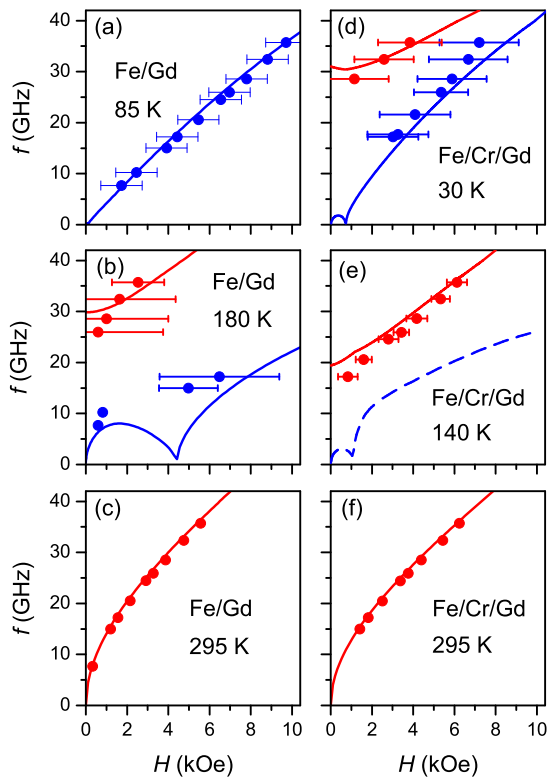}
\caption{Examples of frequency-\textsl{vs}-field dependencies at different temperatures for samples Fe/Gd (a,b,c) and Fe/Cr/Gd (d,e,f). Points are the experimental data, lines are the result of modelling.}
\label{fvsH}
\end{figure}

\begin{figure*}[t]
\includegraphics[width=\textwidth]{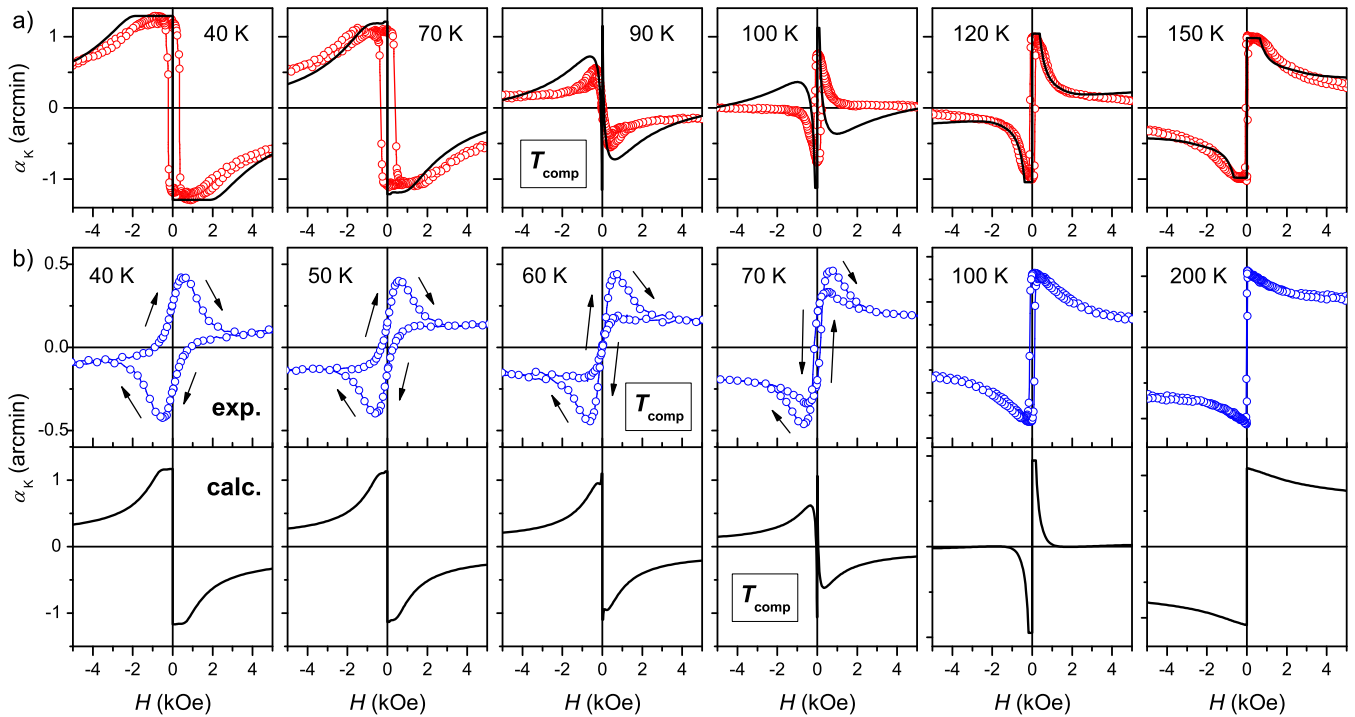}
\caption{MOKE data at different temperatures for Fe/Gd (a) and Fe/Cr/Gd (b) samples. Points are the experimental data, lines are the model calculations.}
\label{Kerr}
\end{figure*}

\begin{figure}[t]
\includegraphics[width=.85\columnwidth]{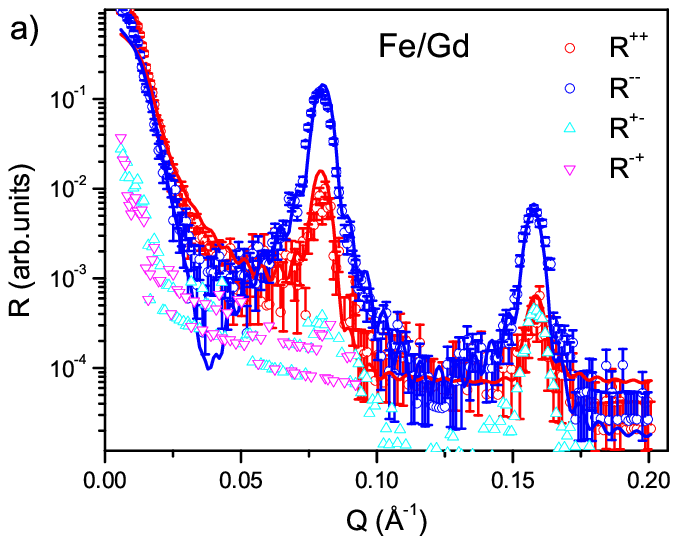}
\includegraphics[width=.85\columnwidth]{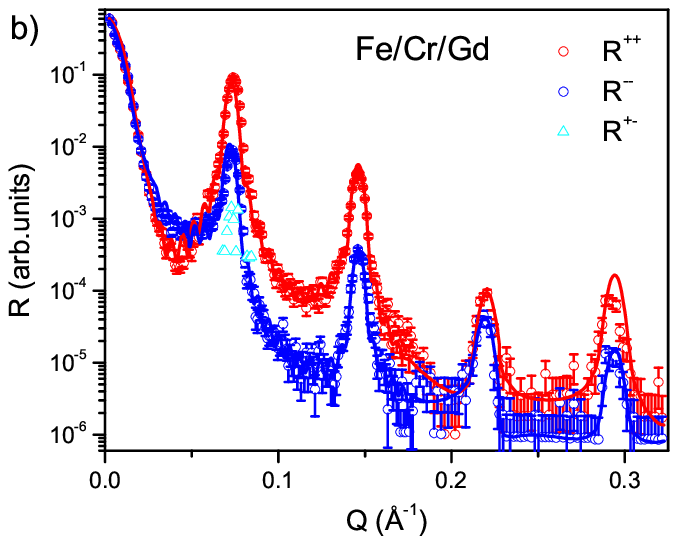}
\caption{Experimental (circles) and fitted (lines) PNR spectra at $T=15$~K, $H=500$~Oe for samples Fe/Gd (a) and Fe/Cr/Gd (b).}
\label{PNR}
\end{figure}

\begin{figure}[t]
\includegraphics[width=.85\columnwidth]{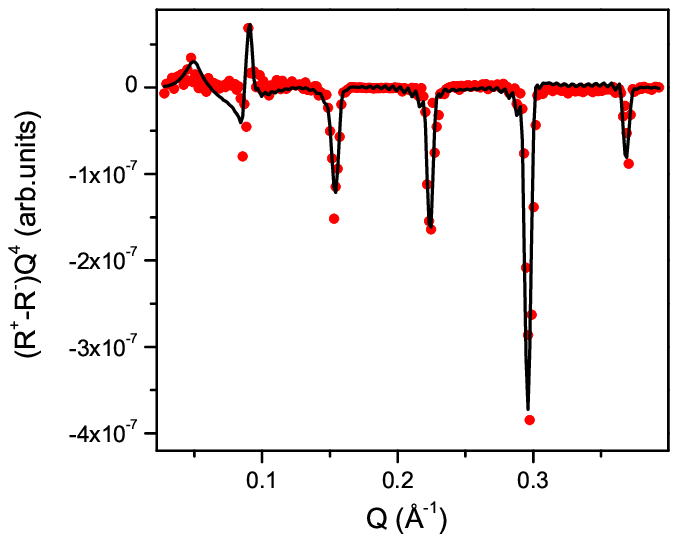}
\caption{Experimental (points) and fitted (lines) RXMR spectrum at $T=15$~K, $H=500$~Oe for the sample Fe/Cr/Gd.}
\label{RXMR}
\end{figure}


Thus, to achieve better agreement between experiment and model, we additionally considered the possibility of varying parameters $M_\mathrm{Fe}$ and $\zeta$ in our fitting procedure. The result of this approach is shown in Fig.~\ref{MH}b by solid lines and the corresponding fitting parameters are presented in Tab.~\ref{Table} (fit~2). As expected, we obtained much better fit of the experimental magnetization curves with increased $M_\mathrm{Fe}$ and decreased parameter $\zeta$.

The inset in Fig.~\ref{MH}b demonstrates experimental and calculated dependencies $m(T)$ for different applied fields. The correspondence between the experiment and theory is good for relatively high field $H=6$~kOe. In the region of low fields, the accordance is not perfect, probably, due to the increasing role of magnetic domain structure. In particular, the model predicts the existence of the compensation point at $T\approx70$~K, while this point is completely obscured by the domain structure in the experimental $m(T)$ curve at $H=300$~Oe.

\subsection{Ferromagnetic resonance}

Magnetic resonance spectra of both studied samples demonstrate two spectral branches (Fig.~\ref{Hres}). One absorption line is observed in the region of high temperatures. At $T=300$~K the resonance peak is relatively narrow ($\Delta H\sim100$~Oe). As temperature decreases, it broadens and shifts towards lower fields. The second peak can be clearly detected at lowest temperatures. However it is more broad and completely disappears at heating.

The resulting temperature dependencies of the resonance fields $H_\mathrm{res}(T)$ are shown in Fig.~\ref{Hres} for frequencies $25.9$~GHz and $35.7$~GHz (examples of experimental spectra are shown in the inset of Fig.~\ref{Hres}a). Note, that the high-field low-temperature peak demonstrates slightly different behaviour for samples Fe/Gd and Fe/Cr/Gd. For the sample Fe/Gd it shifts towards higher fields at heating. On the contrary, for the sample Fe/Cr/Gd it has a tendency to shift towards lower fields.

Examples of frequency \textsl{vs} field dependencies, $f(H)$, at different temperatures are demonstrated in Fig.~\ref{fvsH}. Note that the low-field (high-frequency) mode at $T<200$~K has a gap in the spectrum at $H=0$.

The results of mean-field modelling of FMR are shown by lines in Figs.~\ref{Hres}, \ref{fvsH}. The calculations of eigenfrequencies in the system were performed using model parameters obtained from static magnetization data. For gyromagnetic ratio in Fe and Gd layers, we used the corresponding values for bulk materials: $\gamma_\mathrm{Fe}/2\pi=2.94$~GHz/kOe, $\gamma_\mathrm{Gd}/2\pi=2.80$~GHz/kOe. Following the results of \cite{Drov2017}, we considered the non-local damping term \eqref{Damp} in LLE to suppress the high-order spin-wave modes in Gd which are not observed experimentally. For the parameter of the non-local damping in Gd, we used the value $A^*=0.025$~nm$^2$ estimated in previous work \cite{Drov2017} for the Fe/Gd structure.

In spite of simplicity of the used model, the general correspondence between experiment and theory for both $f(H)$ and $H_\mathrm{res}(T)$ dependencies is reasonable. This fact confirms the applicability of our approach.

Comparing the experimental spectra with model results, the types of precession modes for the observed resonance lines can be identified. The high-field peak observed at low temperatures (Fig.~\ref{Hres}) corresponds to the low-frequency branch of the spectrum (Fig.~\ref{fvsH}~(a,~d)). This mode is associated with in-phase precession of Fe and Gd layers. The line which arises in low fields at higher temperatures corresponds to the high-frequency branch of the spectrum (Fig.~\ref{fvsH}~(b,~e)). This spectral branch demonstrates a gap in the spectrum at $H=0$ and is associated with strongly inhomogeneous ``exchange'' mode. For this mode, the precession phase of the central part of Gd layer is opposite to the precession phase of the Fe layers.

As temperature rises, the gap in the spectrum decreases and the corresponding peak is shifted to higher fields (Fig.~\ref{Hres}). The gap value depends on both exchange stiffness of Gd layers and exchange coupling between Fe and Gd layers. Due to the significant reduction of the interlayer coupling in the Fe/Cr/Gd sample comparing the Fe/Gd structure, it shows much smaller gap in the spectrum at same temperature. Thus, in the spectra measured at fixed frequency, the exchange mode for the sample Fe/Cr/Gd arises at lower temperature (Fig.~\ref{Hres}).

\subsection{Magneto-optical Kerr effect}

Magnetic hysteresis of the samples and its behaviour near the compensation point was investigated by MOKE technique. The penetration depth of the visible light into metal is about $\sim100$~\AA{} \cite{Hahn1995}. Thus, MOKE signal provides information about magnetization in several upper layers of the superlattice. In our experimental geometry the MOKE signal is proportional to the component of magnetization parallel to the applied field. The contribution of the Fe and Gd layers to the total effect (the rotation of the reflected light polarization) is essentially different \cite{Hahn1995}. In particular, it means that the MOKE signal has different sign for Gd- and Fe-aligned phases.

Fig.~\ref{Kerr} demonstrates experimental MOKE hysteresis loops and their comparison with model calculations for different temperatures. To calculate the MOKE signal $\alpha_\mathrm{K}$ from the entire superlattice, we used a simplified approach considering additive contribution of individual layers and exponential decay of the light intensity in the film:
\begin{equation*}
\alpha_\mathrm{K}\sim\int_0^D\alpha(z)M^\parallel(z)e^{-z/\delta}dz,
\end{equation*}
where $D$ is the total thickness of the superlattice, $\delta$ is the penetration depth of the light, $M^\parallel(z)$ is the magnetisation component along the magnetic field as a function of the depth $z$, $\alpha(z)$ is the coefficient which is different for Fe and Gd layers ($\alpha_\mathrm{Fe}$ and $\alpha_\mathrm{Gd}$ respectively).

As it may be seen from Fig.~\ref{Kerr}, the Fe/Gd structure shows relatively narrow hysteresis loops $\lesssim500$~Oe. The compensation temperature $T_\textrm{comp}\approx90$~K can be clearly identified as the point where an inversion of the hysteresis loop occurs. The experimental data can be approximated rather well within the considered model at all temperatures except the region close to the compensation point with parameters $\delta=70$~\AA{} and $\alpha_\mathrm{Fe}/\alpha_\mathrm{Gd}\approx-2$. The plateaus on the MOKE curves in the region of low fields indicate the regions where the collinear phase is realized.

In contrast to the Fe/Gd structure, the MOKE curves for the sample Fe/Cr/Gd demonstrate a strong hysteresis at low temperatures and the correspondence with the calculated curves is not so good. Nevertheless, the experimental loops clearly demonstrate the existence of the compensation at $T\approx60$~K where the remanent MOKE signal turns to zero. At lower temperatures the remanence is negative which can be connected with realization of the Gd-aligned phase. On the contrary, at higher temperatures the remanence is positive indicating the Fe-aligned phase. Note that calculated compensation temperature ($T_\textrm{comp}\approx70$~K) is in reasonably good agreement with the experimental one.

The observed strong low-temperature hysteresis indicates the increasing role of magnetic domain structure in polycrystalline Gd layers for the Fe/Cr/Gd superlattice. As a consequence, the magnetic state in weak fields $\lesssim2$~kOe is strongly dependent on the magnetic history of the sample. Demagnetizing the sample from high fields to $H=0$ initiates the Gd-aligned phase in the system. On the contrary, cooling the sample from high temperatures in a weak field seems not to change the initial Fe-aligned state. Probably, such situation takes place for the static $m(T)$ curve at $H=300$~Oe (Fig.~\ref{MH}) which shows no sign of a minimum at $T_\textrm{comp}$.

\subsection{PNR and RXMR}

The distribution of magnetization within the samples was determined by simultaneous refinement of PNR and RXMR spectra. The neutron and RXMR experimental data at 15~K for 500~Oe magnetic field are displayed in Figs.~\ref{PNR},~\ref{RXMR}. The experiments were performed under the field-cooled conditions. The data analysis involves simultaneous refinement of experimental spectra for polarized neutrons and X-rays as described by E.~Kravtsov, D.~Haskel et.\,al. \cite{Kra2009,Has2012}. The calculation scheme is based on using a unified parameterization of chemical- and element-specific in-plane magnetization profiles in the multilayer. To simplify the calculation, each Gd layer was divided into three sublayers: two interfacial layers and a central layer.

Since there is negligible signal in the spin-flip neutron channel, all the magnetic moments in the systems are aligned along or opposite to the applied magnetic field. The PNR spectra clearly demonstrate different types of magnetic ordering in the samples. For the Fe/Gd superlattice, the Gd-aligned phase is realized, while the Fe/Cr/Gd sample demonstrates the Fe-aligned state.

For both samples, the magnetic moment in Fe layers was found to be close to the bulk value $\approx2.2\mu_\mathrm{B}$, while the magnetization distribution in Gd layers is strongly nonuniform. In the sample Fe/Gd the magnetic moments of Gd layers were found to reach $\approx7\mu_\mathrm{B}$ at interfaces and $\approx5\mu_\mathrm{B}$ in the middle. For the Fe/Cr/Gd structure, the magnetic moment of Gd is $\approx7\mu_\mathrm{B}$ at the Gd/Cr interfaces and $\approx4\mu_\mathrm{B}$ in the middle of the layer (the accuracy is about $0.2\mu_\mathrm{B}$). The interfacial region in Gd is about 10~\AA{} in thickness.

Note that the mean-field model predicts uniform magnetization in Gd under the experimental conditions and does not explain the observed increase of the magnetic moment near the interfaces. Such a ``proximity effect'' seems to be typical for Fe/Gd structures \cite{Hask2001,Choi2004}. Here the same effect is found for the investigated Fe/Cr/Gd superlattice as well.

\section{Discussion}

\begin{figure}[t]
\includegraphics[width=.85\columnwidth]{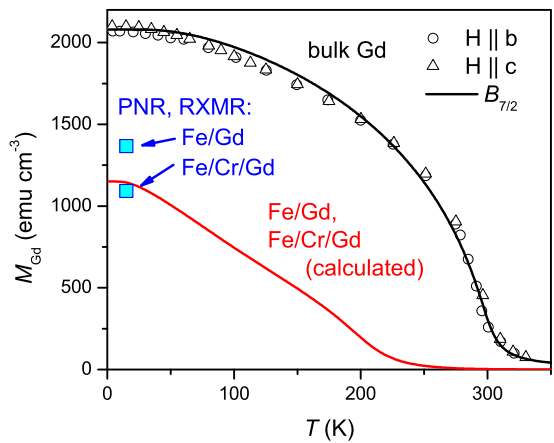}
\caption{Temperature dependence of Gd magnetization in the bulk crystal (experiment and Brillouin function at $H=5$~kOe \cite{Nigh1963}) and in the studied Fe/Gd and Fe/Cr/Gd structures (calculations for $H=0$ in the middle of the Gd layer). The square points are obtained from PNR and RXMR experiments.}
\label{Gd}
\end{figure}

In previous work \cite{Nigh1963}, it was shown that the temperature dependence of magnetization in bulk gadolinium can be described reasonably well by Brillouin function with spin 7/2 (see Fig.~\ref{Gd}). On the contrary, it seems that magnetic properties of thin gadolinium films in Fe/Gd multilayers are poorly described within the standard mean-field model \cite{Drov2017}. Nevertheless, a formal supposition of a temperature dependent mean-field parameter seems to be productive and leads to good approximation of both static and dynamic magnetic properties of the samples. Possible physical arguments for such supposition were discussed in more detail in \cite{Drov2017} where alternative effective field approaches were considered. The comparison of the experiment with the model calculations demonstrates the efficiency of our approach for both Fe/Gd and Fe/Cr/Gd superlattices.

Within the considered model, we can make conclusions about temperature dependence of magnetization $M(T)$ in Gd layers. In particular, the $M(T)$ curve for the central part of Gd layer proves to be close to linear with Curie temperature $T_\mathrm{C}\approx200$~K (which is noticeably lower than the value for bulk Gd, $T_\mathrm{C}\approx290$~K). Note that this result is in a good agreement with \cite{Hos2002jpcm}.

The obtained interlayer AFM exchange energy in the Fe/Gd multilayer is about $J\approx-40$~erg\,cm$^{-2}$. This value recalculated per one interfacial atom gives approximately $J\approx-0.02$~eV~$\approx-200$~K which is in accordance with \cite{Choi2004}. Introduction of the 4~\AA{} thick Cr spacers into Fe/Gd superlattice reduces the interlayer exchange energy by more than an order of magnitude, however the AFM sign of the coupling does not change. Note that the interfacial roughness in the Fe/Cr/Gd multilayer is comparable with the Cr spacer thickness. Thus, we suppose that the observed reduction of the AFM coupling in Fe/Cr/Gd multilayer is due to ``pin-holes'' in the Cr spacer (see also \cite{Drov2015}). In this case, the found coupling constant $J\approx2.5$~erg\,cm$^{-2}$ must be considered as an ``effective'' averaged exchange parameter.

In both Fe/Gd and Fe/Cr/Gd samples, the saturation magnetization values for Fe and Gd layers obtained from the mean-field approximation are noticeably smaller than their bulk values ($M_\mathrm{Fe}^\mathrm{bulk}\approx1750$, $M_\mathrm{Gd}^\mathrm{bulk}\approx2050$~emu\,cm$^{-3}$). In principle, the observed strong reduction of magnetization can be explained by a large degree of structural disorder and amorphousness of the grown FM layers. Indeed, such effects were previously reported for both thin polycrystalline Gd \cite{Rom2011,Ward2013} and amorphous Fe layers \cite{Hand1993}. In both cases the magnetization reduction can reach $\sim50$\% of the bulk value.

On the other hand, the PNR and RXMR data may shed light on another possible mechanism of the observed reduced magnetization in Fe layers. Note that PNR and RXMR confirm the reduction of magnetization in central part of Gd layers. The value $\sim4\mu_\mathrm{B}$ per atom corresponds to approximately $\sim1100$~emu\,cm$^{-3}$ which is in accordance with the mean-field analysis of magnetization data (see Fig.~\ref{Gd}). On the contrary, according to PNR and RXMR, the magnetization of Fe layers is close to the bulk value. However, the Gd magnetization near the Gd-Fe interface is significantly increased up to $\sim7\mu_\mathrm{B}$ due to a ``proximity effect'' and oriented oppositely to the Fe magnetization. The thickness of this region with increased Gd moment is comparable with the interface roughness. The existence of such interface transition region may lead to effective reduction of the net magnetic moment in Fe layers which becomes apparent in magnetization data. A simple estimate shows that one Gd atomic layer magnetized up to saturation value ($\approx7\mu_\mathrm{B}$) is enough to reduce the net magnetization of the Fe layer about 10\%. Thus, the observed reduction ($\approx20-25$\%) is not surprising.

An introduction of the Cr spacer between Fe and Gd seems to suppress the proximity effect initiating the observed increase of the Fe layer magnetization. Another effect of the Cr spacer consists in significant modification of the crystal structure in Gd layers \cite{Kra2017}. The GIXRD spectra demonstrated the coexistence of fcc and hcp Gd crystal phases in the Fe/Cr/Gd multilayer while the Fe/Gd structure showed only the presence of hcp Gd phase. Magnetic studies demonstrated an increasing role of domain structure in the Fe/Cr/Gd sample as compared to the Fe/Gd structure. At the same time, the mean-field analysis of the magnetization data showed a noticeable change of the parameter $\zeta$. This result seems logical because $\zeta$ can be considered as a direct parameter of the crystal structure. However, due to polycrystalline structure of the real layers this parameter has only effective character. On the other hand, according to Eq.~\eqref{ExchStiff}, it has a direct connection with the exchange stiffness of Gd layers. The resulting calculated low-temperature values of the exchange stiffness in Gd layers are $A=0.75\cdot10^{-7}$~erg\,cm$^{-1}$ for Fe/Gd and $A=0.57\cdot10^{-7}$~erg\,cm$^{-1}$ for Fe/Cr/Gd structure.

Note that a previous study of Fe/Cr/Gd structures \cite{Drov2015} neglected to account for magnetization twist states in the Gd layers. For this reason, an additional biquadratic term in the interlayer exchange energy was considered in \cite{Drov2015} for better description of the experimental data, in particular, to explain the strongly non-linear $M(H)$ curves at low temperature. More detailed data obtained in the present work demonstrated the important role of inhomogeneous magnetization distribution in the Gd layers for both Fe/Gd and Fe/Cr/Gd structures. Using the developed mean-field approach, we achieved reasonable agreement between the experiment and the model simulations considering only the usual Heisenberg-type exchange at the interface between FM layers.

\section{Conclusion}

In this work, we performed comparative studies of structural and magnetic properties of [Fe/Gd]$_{12}$ and [Fe/Cr/Gd/Cr]$_{12}$ superlattices. The experimentally obtained magnetization curves and FMR spectra were analysed in the frame of mean-field approximation in the wide range of temperatures $4-300$~K using the modified approach of the work \cite{Drov2017} which takes into account the temperature dependence of the mean-field parameter in Gd layers. We confirm that this approach provides reasonably good correspondence between the experimental data and model calculations for both samples.

The performed model calculations allowed us to obtain magnetic parameters of the Fe/Gd and Fe/Cr/Gd superlattices and analyse the influence of the Cr spacer on their magnetic properties. The main effect of the Cr spacers introduced in the Fe/Gd superlattice is a strong reduction of the exchange coupling between Fe and Gd layers. At the same time we also observe modification of magnetic properties of both FM layers which can be connected with their structural changes.

For both investigated samples the FM layers have reduced values of saturation magnetization as compared to the bulk Fe and Gd. This effect can be explained by large degree of structural disorder and amorphousness of the grown FM layers as well as by imperfections of the interfaces leading to existence of a transition layer with reduced magnetization due to a strong AFM coupling between Fe and Gd atoms (``proximity effect''). The PNR and RXMR experiments clearly demonstrated the existence of such a transition layer with strongly increased Gd magnetization. The introduction of Cr spacers between Fe and Gd layers seems to suppress this effect initiating a slight increase of the net magnetization in Fe layers. At the same time we observe the decrease of exchange stiffness of Gd layers in Fe/Cr/Gd structure and increasing role of magnetic domains. These effects seem to be connected with formation of fcc crystallites in Gd layers.

\section*{Acknowledgements}

This work is based upon experiments performed at the NREX instrument operated by the Max-Planck Society at the Heinz Maier-Leibnitz Zentrum (MLZ), Garching, Germany.

Work at APS is supported by the U.S. Department of Energy (DOE), Office of Science, under Contract No. DE-AC02-06CH11357.

Research in Ekaterinburg was performed  in terms of the state assignment of Federal Agency of Scientific Organizations of the Russian Federation (theme ``Spin'' No.~AAAA-A18-188020290104-2). X-ray measurements were performed at the Collective Use Center of IMP.

The work is partially supported by the Russian Foundation for Basic Research (grants No.\,16-02-00061, No.\,18-37-00182) and by the Ministry of Education and Science of the Russian Federation (grant No.\,14-Z-50.31.0025).

We would like to thank A.~Mukhin, V.~Ivanov and A.~Kuz'menko (GPI~RAS) for assistance in performing measurements on a SQUID magnetometer.

\end{document}